\documentclass[prb,twocolumn,showpacs,preprintnumbers,amsmath,amssymb]{revtex4}

\usepackage[dvips]{graphicx}
\usepackage{color}
\usepackage{dcolumn}
\usepackage{bm}

\begin{document}

\title{Finite size scaling of the de Almeida-Thouless instability in random sparse networks}

\author{Hisanao Takahashi$^1$, Federico Ricci-Tersenghi$^2$ and Yoshiyuki Kabashima$^1$}

\affiliation{
$^1$Department of Computational Intelligence and Systems Science, Tokyo Institute of Technology, Yokohama 226-8502, Japan \\
$^2$Dipartimento di Fisica, Sapienza Universit\`a di Roma, INFN, Sezione di Roma I, 
IPCF - CNR, P.le Aldo Moro 2, I-00185 Roma, Italy}

\date{\today}

\begin{abstract}
We study, in random sparse networks, finite size scaling of the spin glass susceptibility $\chi_{\rm SG}$, which is a proper measure of the de Almeida-Thouless (AT) instability of spin glass systems. Using a phenomenological argument regarding the band edge behavior of the Hessian eigenvalue distribution, we discuss how $\chi_{\rm SG}$ is 
evaluated in infinitely large random sparse networks, which are usually identified with Bethe trees, and 
how it should be corrected in finite systems. In the high temperature region, data of extensive numerical experiments are generally in good agreement with the theoretical values of $\chi_{\rm SG}$ determined from the Bethe tree. In the absence of external fields, the data also show a scaling relation $\chi_{\rm SG}=N^{1/3}F(N^{1/3}|T-T_c|/T_c)$, which has been conjectured in the literature, where $T_c$ is the critical temperature. In the presence of external fields, on the other hand, the numerical data are not consistent with this scaling relation. A numerical analysis of Hessian eigenvalues implies that strong finite size corrections of the lower band edge of the eigenvalue distribution, which seem relevant only in the presence of the fields, are a major source of inconsistency. This may be related to the known difficulty in using only numerical methods to detect the AT instability. 
\end{abstract}


\maketitle

\section{Introduction} 
The discovery of spontaneous replica symmetry breaking (RSB) in the low temperature region of spin glass (SG) models \cite{Giorgio79} is perhaps the most significant achievement in the field of statistical mechanics of disordered systems. It was first found in the analysis of the Sherrington-Kirkpatrick (SK) model, which is a fully connected mean field SG model, by introducing the Parisi ansatz for describing an RSB state as a relevant saddle point of the replicated free energy, which is intrinsically replica symmetric. The mathematical technicalities of constructing the solution generated tremendous controversy in the early days of SG research. However, alternative mean field approaches \cite{TAP,cavity} and mathematically rigorous arguments \cite{Talagrand,inequality} now support the correctness of the Parisi solution. These days, it is commonly accepted that RSB does occur in a class of mean field models although the existence of RSB in systems of finite dimension is still under debate \cite{debates}.

In the present situation, SG models on random sparse networks play a special role. A random sparse network is constructed such that each node is randomly coupled to a finite number of other nodes. As the random construction guarantees statistical uniformity of the network structure, the random sparse model is classified as a mean field model that exhibits RSB. At the same time, unlike fully connected models, a concept of adjacency between nodes is naturally introduced in the random sparse network, which may make it possible to characterize the RSB transition by using the concept of correlation length (as in the standard analysis of models in finite dimensions). For very large (possibly infinitely large) random sparse networks, a solution for the SG model exists whenever the SG correlation decays fast enough (in practice, in the replica symmetric phases). This solution is obtained by assuming equivalence between the random sparse network and the corresponding Bethe tree.

In SG models that undergo a continuous phase transition as the temperature decreases, the onset of RSB is signaled by the de Almeida-Thouless (AT) instability \cite{dAT}, which corresponds to the divergence of the SG susceptibility. The AT instability naturally defines a critical line $T_c(H)$ in the temperature ($T$) -- external field ($H$) plane. The shape of this line in the fully connected SK model is very peculiar: $H_c(T)$ diverges for $T \to 0$; that is, an RSB phase exists for any value of the external field. This behavior is rather unusual and can take place only in models where the coordination number diverges in the thermodynamic limit. In more realistic situations, $T_c(H)$ should become 0 when the external field reaches a critical value $H_c$: SG models on random sparse networks display this more realistic behavior.

A very interesting (and still largely open) question is how the AT instability is observed in finite systems. Numerical studies of the SK model have failed to identify the critical point 
with $H \neq 0$ and have reported very strong finite size effects
\cite{Aspelmeier}. For SG models on random sparse networks, much less
is known: there are {
very few numerical studies \cite{Jorg}} and a validation of the finite size scaling relations is still lacking.

In this paper, we mainly examine how the critical condition for the AT instability obtained in the thermodynamic limit should be corrected for finite sized systems.

A proper measure for detecting the AT instability is the spin glass susceptibility defined as~\footnote{Our working definition of $\chi_{\rm SG}$ differs from the standard one by a factor $T^2$, which is, however, irrelevant in the study of its divergence.} 
\[ \chi_{\rm SG}=N^{-1}\sum_{i,j} \overline{\left (\left \langle S_i S_j \right \rangle -\left \langle S_i \right \rangle \left \langle S_j \right \rangle \right)^2}\;, \]
where $N$ is the number of spins, $S_i$ is a spin variable, and $\langle \cdots \rangle$ (respectively, $\overline{\cdots}$) denotes thermal (respectively, configurational or disorder) averages. 
For $T>T_c(H)$, in the infinitely large system limit, a random sparse network can be accurately approximated by a Bethe tree, thus providing a direct expression of $\chi_{\rm SG}$. On the other hand, for large but finite systems, that expression for $\chi_{\rm SG}$ has to be corrected in an appropriate manner to account for finite size effects. We will show that the edge behavior of the eigenvalue distribution $\rho(\lambda)$ of the susceptibility matrix plays a key role in this correction. That is, when $\rho(\lambda) \propto (\lambda-\lambda_{\rm min})^\gamma$ holds in the vicinity of the lower band edge of the distribution, i.e., near $\lambda_{\rm min}$, the finite size correction amounts to a scaling relation $\chi_{\rm SG}=N^{\omega} F(t N^{\omega})$, where $\omega=(1-\gamma)/(1+\gamma)$ and $t=(T-T_c)/T_c$, unless $\lambda_{\rm min}$ is significantly influenced by finite size effects. 
In the absence of external fields, the results of random matrix theory indicate $\gamma=1/2$, which yields a known relation for mean field SG models $\chi_{\rm SG}=N^{1/3}F(t N^{1/3})$. Data obtained from extensive numerical experiments confirm that this relation is fairly accurate. On the other hand, numerical data for $N\le 2^{10}$ show a considerable discrepancy with the same scaling relation in the presence of external fields even when $\omega$ is optimally tuned.
Numerical analysis on the Hessian for relatively smaller systems of $N \le 2^8$ indicates that $\lambda_{\rm min}$ has strong finite size corrections in the presence of fields, whereas the profile near $\lambda_{\rm min}$ of the eigenvalue distribution does not change much, which makes it practically difficult to identify $T_c$ by using only numerical methods. This may be the reason why the AT instability is hard to observe in finite-dimensional models.

This paper is organized as follows. The next section introduces the model to be examined. In section 3, we show how $\chi_{\rm SG}$ is evaluated for infinitely large random sparse networks. We also derive the finite size scaling relation on the basis of phenomenological considerations about the eigenvalue distribution of the susceptibility matrix. In section 4, we discuss the numerical experiments examining the validity of the results obtained in section 3. The final section is devoted to a summary.

\section{Model definition}
We will study SG models defined on $C$-regular random graphs in the presence of an external field $H$. The Hamiltonian is given by 
\begin{equation}
{\cal H}(\mbox{\boldmath$S$}) = -\sum_{(ij)\in E} J_{ij} S_i S_j - H \sum_i S_i\;, 
\label{Ham} 
\end{equation}
where $E$ is the set of edges in the random graph, which is chosen uniformly among all graphs of $N$ nodes and $M=N C/2$ edges, having exactly $C$ edges per node. Each node contains an Ising spin $S_i \in \{+1,-1\}$ ($i=1,\ldots,N$), and the couplings are quenched i.i.d.\ random variables extracted from 
\begin{eqnarray}
P(J) = \frac{1}{2} \Big[\delta (J-1)+\delta(J+1) \Big]\;. 
\label{P(J)=1/2...} 
\end{eqnarray}
As already mentioned, we denote the thermal averages with respect to the canonical distribution of inverse temperature $\beta=T^{-1}$ as $\left \langle \cdots \right \rangle =\sum_{\mbox{\boldmath$S$}} (\cdots) \exp \left [-\beta {\cal H}(\mbox{\boldmath$S$}) \right ]/Z(\beta)$, where $Z(\beta)=\sum_{\mbox{\boldmath$S$}} \exp \left [-\beta {\cal H}(\mbox{\boldmath$S$}) \right ]$ is the partition function. Configurational averages with respect to the generation of the couplings and of the graph are denoted as $\overline{(\cdots)}$.

\section{de Almeida-Thouless instability} 
\subsection{Infinitely large systems}
Thermal averages are difficult to evaluate computationally. However, when a given graph is free of cycles (i.e., it is tree), it is known that the Bethe approximation is exact and the belief propagation (BP) algorithm provides exact averages in a practical time \cite{Pearl,KabashimaSaad,MezardParisi}. The BP iterative method is as follows: 
\begin{equation}
\label{u_mu to l t+1}
u_{i \to j}^{\tau+1} = \tanh^{-1} \bigg[\tanh (\beta J_{ij}) \tanh\bigg(\beta H + \sum_{k \in \partial i \setminus j} u_{k \to i}^\tau\bigg)\bigg]\;, 
\end{equation}
where $u_{i \to j}^\tau$ represent message variables which are transmitted between nodes, $\partial i$ is the set of neighbors of $i$, and $\backslash j$ stands for exclusion of $j$. From the fixed point of Eq.~(\ref{u_mu to l t+1}), the thermal average of spin $S_i$ is evaluated as 
\begin{equation}
m_i= \left \langle S_i \right \rangle 
=\tanh \bigg[\beta H+\sum_{j \in \partial i} u_{j \to i}\bigg]\;. 
\label{correct_average} 
\end{equation}

The results obtained by the BP algorithm, Eq.~(\ref{u_mu to l t+1}), are just an approximation for general lattices with loops. However, the length of cycles in a random sparse network typically grows as $O(\ln N)$ as the size of the graph $N$ tends to infinity. This implies that for high temperatures, $T > T_c$, where the spatial correlations decay fast enough, thermal averages in sufficiently large random networks can be accurately evaluated by using BP, i.e.\ as if they were computed on a tree. This idea is also useful for analyzing the AT instability.

Let us focus on a spin $S_i$ in a large random sparse network and approximate the lattice with a tree rooted at $i$. A distinctive property of the tree is that any pair of points is linked by a unique path. Noticing this, we can define a path connecting $i$ with $j$, placed at distance $G$ from $i$, and assign a label $g=0,1,\ldots,G$ to the nodes along the path from $i$ to $j$ ($g=0$ and $g=G$ correspond to $i$ and $j$, respectively). On this tree, the two-point correlation between $i$ and $j$ can be exactly computed as 
{
\begin{eqnarray}
&& \langle S_i S_j \rangle - \langle S_i \rangle \langle S_j \rangle = \beta^{-1} \frac{\partial m_{i}}{\partial H_j} = \nonumber \\
&& = \beta^{-1} \frac{\partial m_{i}}{\partial u_{1 \to 0}} \cdot \frac{\partial u_{1 \to 0}}{\partial u_{2 \to 1}} \cdots \frac{\partial u_{G \to G-1}}{\partial H_j} = \nonumber \\
&& = 
\frac{\partial m_{i}}{\partial u_{1 \to 0}} \prod_{g=1}^G \frac{\partial u_{g \to g-1}}{\partial u_{g+1 \to g}}\;. 
\label{two_point} 
\end{eqnarray}
}%
Here, the derivative with respect to the field $H_j$ acting on $j$ has been replaced by the one with respect to any BP message arriving at site $j$. Statistical uniformity guarantees that the configurational average of the square of Eq. (\ref{two_point}) 
{
\[ \overline{(\langle S_i S_j \rangle - \langle S_i \rangle \langle S_j \rangle)^2} = 
\overline{\left(\frac{\partial m_i}{\partial u_{1 \to 0}} \right)^2 \prod_{g=1}^G \left(\frac{\partial u_{g \to g-1}}{\partial u_{g+1 \to g}} \right)^2}, \]
}%
depends only on $G$. On the other hand, the number of spins at a distance $G$ from $i$ is given by $C(C-1)^{G-1}$ on the tree. Thus, the spin glass susceptibility $\chi_{\rm SG}$ can be evaluated as 
{
\begin{eqnarray}
\chi_{\rm SG}&=&\overline{\left(1-m_i^2\right)^2} \cr 
&\phantom{=}& +\sum_{G=1}^\infty 
C(C-1)^{G-1} \overline{\left (\left \langle S_0 S_G \right \rangle - \left\langle S_0 \right\rangle \left\langle S_G \right \rangle \right)^2}\cr &\propto & \frac{1}{1-(C-1) e^{-\Psi}}, 
\label{chisg} 
\end{eqnarray}
}%
for $T$ approaching $T_c$ from above. The quantity $\Psi$ is the inverse of the SG correlation length
\begin{eqnarray}
\Psi&=&-\lim_{G \to \infty} \frac{1}{G}\ln \overline{\left (\left \langle S_0 S_G \right \rangle - \left\langle S_0 \right\rangle \left\langle S_G \right \rangle \right)^2} \cr &=&- \lim_{G \to \infty} \frac{1}{G} \ln \overline{\prod_{g=1}^G \left (\frac{\partial u_{g \to g-1}}{\partial u_{g+1 \to g}} \right)^2}. 
\label{psi} 
\end{eqnarray}
From Eq. (\ref{chisg}), we obtain the condition for the divergence of $\chi_{\rm SG}$ 
\begin{eqnarray}
(C-1)e^{-\Psi}=1\;,
\label{ATcondition}
\end{eqnarray}
giving the AT instability for infinitely large systems \cite{Rivoire2004,Krzakala}.

Three points are noteworthy. First, unlike the usual critical phenomena in finite dimensions, the AT instability on random sparse networks is not accompanied by the divergence of the correlation length: even at the critical point, the correlation length is finite and equal to $\xi=\Psi^{-1}=1/\ln (C-1)$. This is because the number of nodes at a distance $G$ from a fixed node in a random network grows exponentially fast and proportionally to $(C-1)^G$, which is not the case in models of finite dimensions. Second, for $H=0$, we can obtain an analytical expression for $\chi_{\rm SG}$. Indeed, as long as $(C-1)\tanh(\beta)^2 < 1$, a unique convergent solution of Eq.~(\ref{u_mu to l t+1}) exists and is given by $u_{i \to j}=0$, which in turn implies $\overline{\left (\left \langle S_0 S_G \right \rangle - \left\langle S_0 \right\rangle \left\langle S_G \right \rangle \right)^2} =[\tanh(\beta)]^{2G}$ and 
\begin{eqnarray}
\chi_{\rm SG}&=&\frac{1+\tanh(\beta)^2}{1-(C-1)\tanh(\beta)^2}\;. 
\label{ATdivergence} 
\end{eqnarray}
The above equation yields the critical temperature by setting $(C-1)\tanh(\beta_c)^2=1$, and it shows that as $T \to T_c$ from above, $\chi_{\rm SG}$ diverges as $O(|t|^{-1})$, where $t=(T-T_c)/T_c$, in perfect agreement with the known AT instability in the absence of an external field \cite{Thouless,MP87}. Third, one can still numerically assess Eq.~(\ref{ATcondition}) in a practical time even in the presence of an external field. For this, we utilize a property of BP operating on a Bethe tree whereby the distributions of the message variables for typical sample systems can be obtained as a set of solutions of functional equations, 
\begin{equation}
\pi(u) = \int\prod_{\mu=1}^{C-1} du_\mu\,\pi(u_\mu)\; \overline{\delta\bigg(u- f\Big(\beta J, \beta H + \sum_{\mu=1}^{C-1} u_\mu\Big)\bigg)}, 
\label{hat{pi}^t+1(u)=} 
\end{equation}
where $f(x,y) \equiv \tanh^{-1}\big(\tanh(x)\tanh(y)\big)$. Eq.~(\ref{hat{pi}^t+1(u)=}) can be solved numerically with a sampling method 
in a reasonable time \cite{MezardParisi}. This implies that, for given $G$, a sample of $\left \langle S_0 S_G \right \rangle -\left \langle S_0 \right \rangle \left \langle S_G \right \rangle$ can be generated by Eq.~(\ref{two_point}) from the following BP dynamics: 
\begin{equation}
\label{u_g-1=1/beta...} u_{g \to g-1} = f(\beta J_g,\;u_{g+1 \to g} + r_g) 
\end{equation}
where $g=G,G-1,\ldots,1$, and $J_g$ and $ r_g $ represent independent random numbers respectively sampled from Eq.~(\ref{P(J)=1/2...}) and from the distribution 
\begin{eqnarray}
\mu(r) =\int\prod_{\mu=1}^{C-2} du_\mu \,\pi(u_\mu)\; \delta\left(r-\beta H -\sum_{\mu=1}^{C-2}u_\mu\right)\;. 
\label{rand_merge} 
\end{eqnarray}
Here, $r_g$ stands for the sum of messages from $C-2$ branches that merge with the $g$-th node on the path. The computational cost of this evaluation scales as $O(G)$ per sample, which is computationally feasible. Therefore, for fixed $G$, one can numerically evaluate $\overline{\left (\left \langle S_0 S_G \right \rangle - \left\langle S_0 \right\rangle \left\langle S_G \right \rangle \right)^2}$ by using the sampling stochastic process of Eqs.~(\ref{u_g-1=1/beta...})--(\ref{rand_merge}) many times. Estimates for several $G$ can be then interpolated with an appropriate polynomial in $1/G$. This makes it possible to practically compute Eq.~(\ref{psi}) by extrapolation of $1/G \to 0$ in the fitted polynomial. Table \ref{Tc} lists estimates of $T_c$ for $H=0, 0.1, 0.2$ and $0.3$ for the case of $C=4$. The values for $H \ne 0$ were evaluated by extrapolating fourth degree polynomials fitted to data of $G=1,2,\ldots,20$ that were computed from $10^7$ samplings. $T_c=1.5187\ldots$ for $H=0$ can be obtained by solving $3 \tanh^2(1/T_c)=1$. $\Psi$ varies linearly with respect to $t$ around the critical temperature $T_c$. This implies that $\chi_{\rm SG}$ scales as $O(t^{-1})$ close to $T_c$ in the limit of $N \to \infty$.

\begin{table} 
\begin{tabular}{c|cccc} 
\hline
\hline ~$H$~ & 0.0 & 0.1 & 0.2 & 0.3 \\
\hline ~$T_c $~ & 1.51865 & 1.3053 & 1.1808& 1.0770 \\
\hline 
\end{tabular} 
\caption[]{The critical temperatures $T_c$ for fixed connectivity 4 and four different external fields $H$.} 
\label{Tc} 
\end{table}

We also checked that the method for computing the AT line is equivalent to the method based on perturbing the BP messages and then observing the subsequent evolution of the perturbation, the critical temperature being defined as the lowest temperature such that the perturbation does not grow under BP iteration~\cite{PPR03}.

\begin{figure} 
\begin{center} 
\includegraphics[width=\columnwidth]{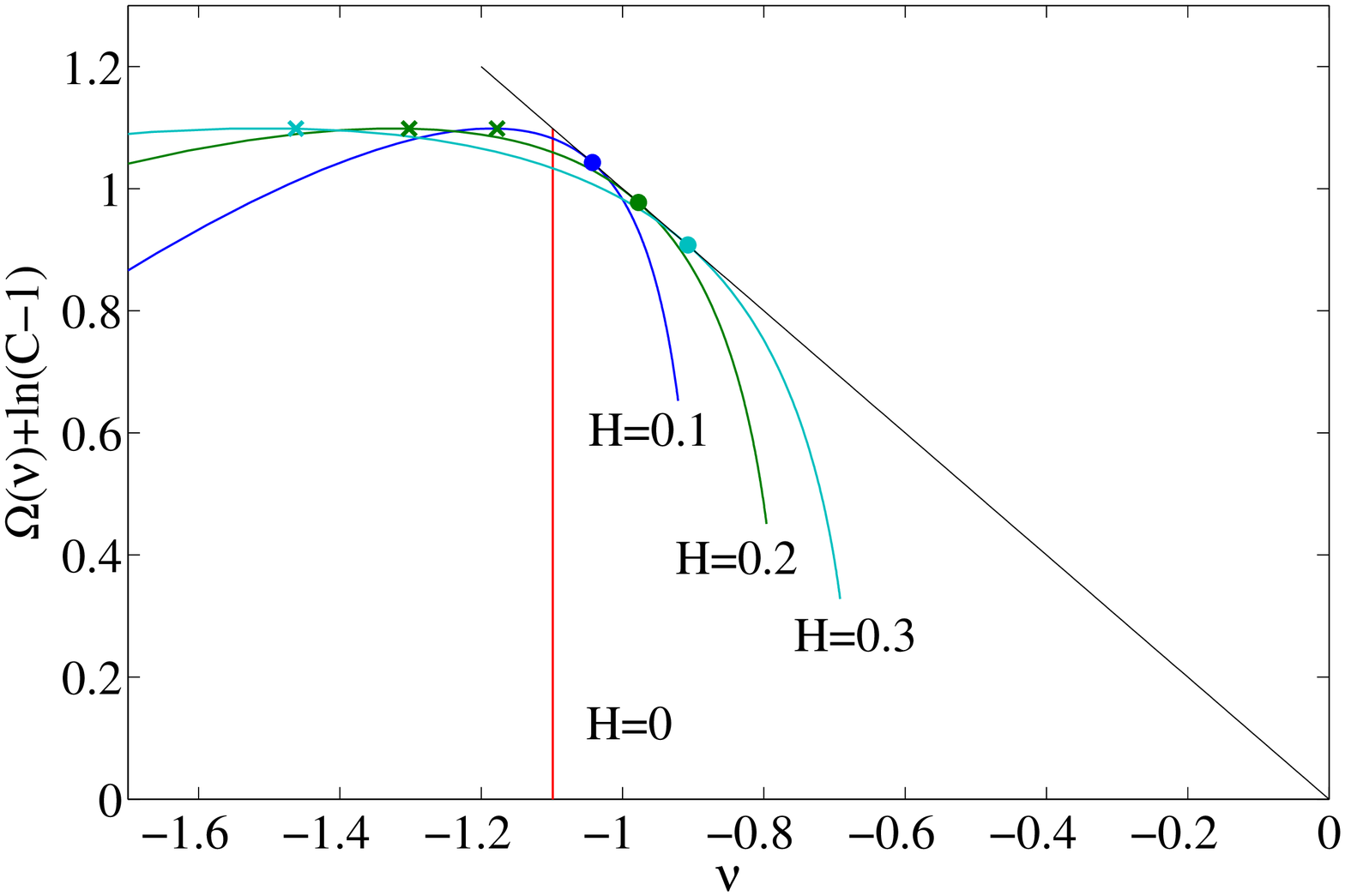} 
\end{center} 
\caption{Profiles of rate function $\Omega(\nu)$ at the AT criticality for several values of external field $H$ in the case of $C=4$. Values of the critical temperatures are shown in table \ref{Tc}. In order to evaluate $\Omega(\nu)$, we first computed $\Psi(s)=-\lim_{G \to \infty}(1/G)\ln \overline{\left (\left \langle S_0 S_G \right \rangle - \left\langle S_0 \right\rangle \left\langle S_G \right \rangle \right)^{2s}}$ by extrapolating numerical data for $G=1,2,\ldots,20$ to $G \to \infty$. Applying a Legendre transformation to this function yields $\Omega(\nu)$ as follows: $\nu=-(\partial/\partial s)\Psi(s)$ and $\Omega= -s \nu -\Psi(s)$, where $\Omega$ has been parameterized by a conjugate variable $s$. For drawing the profiles shown in the figure, we numerically evaluated $\overline{\left (\left \langle S_0 S_G \right \rangle - \left\langle S_0 \right\rangle \left\langle S_G \right \rangle \right)^{2s}}$ based on $10^7$ samples of eq. (\ref{u_g-1=1/beta...}) and varied $s$ in the range of $-1 \le s \le 9$. The profiles for $H \neq 0$ indicate that the dominant values of $\nu$ for the AT criticality (dots) are considerably larger than the most probable values of $\nu$ (crosses). The physical implication of this is that the AT instability for $H \ne 0$ is induced by a small number of atypically large spin correlations.} 
\label{Omega} 
\end{figure}

Note that although the graph is regular, i.e.\ all the vertices are equivalent, the presence of a uniform field produces many heterogeneities and does not allow for the existence of a factorized solution (and this may eventually lead to larger fluctuations in finite systems \cite{ParisiRizzo09}) . The main effect is that the two-spin correlation $\langle S_i S_j \rangle - \langle S_i \rangle \langle S_j \rangle$ fluctuates a lot between different pairs of spins separated by the same distance. Figure~\ref{Omega} plots the rate function $\Omega(\nu)=G^{-1} \ln P(\nu)$ of the probability distribution $P(\nu)$ for the logarithm of the correlation  
\[ \nu = \lim_{G \to \infty} \frac1G \ln \Big[ (\langle S_0 S_G \rangle - \langle S_0 \rangle \langle S_G \rangle)^2 \Big]\;. \]
Correlations contributing the most to the assessment of $\Psi = -\text{argmax}\Big[ \Omega(\nu) + \nu \Big]$ are marked by dots in Fig.~\ref{Omega}, and are clearly larger than the most probable correlations corresponding to the maximum of $\Omega(\nu)$. This means that, as soon as $H \neq 0$, the long-range order in the model is produced essentially by very few pairs of strongly correlated spins, while the vast majority of pairs of spins remain uncorrelated.

\subsection{Finite systems} 
\label{finiteSize} 
So far, we have reviewed how $T_c$ for the AT instability can be evaluated by computing $\chi_{\rm SG}$ in infinitely large random sparse networks. However, $\chi_{\rm SG}$ is intrinsically upper-bounded by $N$ and never diverges as long as $N$ is finite. Therefore, we have to examine how the scenario in the previous section should be modified in finite systems so that we can appropriately analyze data from numerical experiments.

A naive correction taking the finiteness of the system into account can be made by truncating the summation of Eq. (\ref{chisg}) at a finite number $G=G_{\rm max}$, which leads to 
\begin{eqnarray}
\chi_{\rm SG} \propto \frac{1-((C-1)e^{-\Psi})^{G_{\rm max}}}{1-(C-1)e^{-\Psi}}, 
\label{wrong_correction} 
\end{eqnarray}
since $G$ cannot tend to infinity when the system is finite. Unfortunately, the following considerations indicate that such a correction is not appropriate for describing the behavior in the vicinity of $T_c$. The right-hand side of Eq.~(\ref{wrong_correction}) gives $G_{\rm max}$ when the critical condition $1-(C-1)e^{-\Psi} \to 0$ holds. $G_{\rm max}$ grows monotonically as $N$ increases. However, the growth rate is only $O(\ln N)$ since $N \sim C(C-1)^{G_{\rm max}}$ must hold for satisfying the constraint concerning the number of nodes. This rate is obviously too slow since numerical experiments show that the $\chi_{\rm SG}$ of finite systems grows as $O(N^{1/3})$ at the critical condition, at least, for $H=0$ \cite{Jorg}.

This discrepancy indicates that effects of self-inter\-actions, which are ignored in the Bethe tree approximation, must be taken into account when evaluating the dependence of $\chi_{\rm SG}$ on the system size $N$ in the vicinity of $T_c$. Unfortunately, such an evaluation requires a complicated calculation and still does not lead to an accurate expression in general. Therefore, to avoid these technical difficulties, we shall employ a phenomenological derivation.

Consider the Hessian $A=\chi^{-1}$, where $\chi$ denotes a susceptibility matrix with elements $\chi_{ij}=(\left \langle S_i S_j \right \rangle- \left \langle S_i \right \rangle \left \langle S_j \right \rangle)$. As a working hypothesis, we assume that the eigenvalues of $A$, $\lambda_1\le \lambda_2 \le \ldots \le \lambda_N$, obey a continuous distribution $\rho(\lambda)$, which behaves as 
\begin{eqnarray}
\rho(\lambda) \propto (\lambda-\lambda_{\rm min})^\gamma\;, 
\label{band_edge} 
\end{eqnarray}
close to the lower band edge $\lambda_{\rm min}$ for $T > T_c$ and $N \to \infty$. Moreover, we shall assume for the moment that $\lambda_{\rm min}$ is not heavily modified by finite corrections. For $H=0$, an analysis of random matrices of fixed weights, in conjunction with Thouless-Anderson-Palmer theory \cite{TAP}, implies that the distribution can be expressed as 
\begin{eqnarray}
\rho(\lambda;\Lambda,\mu)= \frac{1}{2\pi} \frac{\sqrt{4(C-1)\Lambda^2-(\lambda-\mu)^2}}{C\Lambda^2-(\lambda-\mu)^2/C}, 
\label{eigenvalue} 
\end{eqnarray}
using certain parameters $\Lambda$ and $\mu$ \cite{sparse_distribution}, which supports $\gamma=1/2$ for $C \ge 3$. However, here, we do not exclude the possibility that $\gamma$ may depend on $T$ and $H$. Equation~(\ref{band_edge}) provides another expression of $\chi_{\rm SG}$ 
\begin{eqnarray}
\chi_{\rm SG}=\frac{1}{N} \sum_{k=1}^N \overline{\lambda_k^{-2}} \to \int_{\lambda_{\rm min}} d\lambda\,\lambda^{-2} \rho(\lambda) \propto \lambda_{\rm min}^{-(1-\gamma)}, 
\label{another_chi_SG} 
\end{eqnarray}
as $N \to \infty$. Assuming that $\chi_{\rm SG}$ diverges as $O(|t|^{-1})$ at criticality, $\lambda_{\rm min} \propto t^{1/(1-\gamma)}$ holds as $T$ approaches $T_c$ from above in the limit of $N \to \infty$.

However, the statistical fluctuations of the eigenvalues are not negligible around $T_c$ for large but finite $N$. As a first approximation, therefore, let us regard $\lambda_k$ $(k=1,2,\ldots,N)$ as independently and identically distributed (i.i.d.) random variables extracted from $\rho(\lambda)$. Since $\lambda_1$ is the smallest value among the $N$ i.i.d. random variables, the theory of extreme value statistics \cite{Gumbel} indicates that magnitude of the fluctuation of $\lambda_1$ can be evaluated by a simple equation: 
\begin{eqnarray}
N \int_{\lambda_{\rm min}}^{\lambda_1} d\lambda\, \rho(\lambda) \sim O(1), 
\label{lambda1fluct} 
\end{eqnarray}
which yields a scaling relation $\lambda_1-\lambda_{\rm min} \propto N^{-1/(1+\gamma)}$.

Replacing $\lambda_{\rm min}$ by its scaling relation in terms of $t^{1/(1-\gamma)}$, Eq.~(\ref{lambda1fluct}) leads to the following expression for the smallest eigenvalue: 
\begin{eqnarray}
\lambda_1 &=& A_1 t^{1/(1-\gamma)} + N^{-1/(1+\gamma)} \xi_1 = \cr &=& N^{-1/(1+\gamma)} \left (A_1 (tN^{\frac{1-\gamma}{1+\gamma}})^{1/(1-\gamma)}+ \xi_1 \right), 
\label{lambda1fluct2} 
\end{eqnarray}
where $A_1$ is a constant and $\xi_1$ is a random variable taking values $O(1)$. This derivation also indicates that $\lambda_k$ can be expressed similarly to Eq.~(\ref{lambda1fluct2}) as long as $k \sim O(1)$. Accordingly, all contributions to $\chi_{\rm SG}$ from $\lambda_k$ with $k \sim O(1)$ can be summed together in a scaling relation like 
\[ \frac{1}{N}\sum_{k\sim O(1)} \overline{\lambda_k^{-2}} \sim N^{\frac{1-\gamma}{1+\gamma}}g\left(t N^{\frac{1-\gamma}{1+\gamma}}\right)\;, \]
after being averaged with respect to the $\xi_k$. Here, $g(x)$ is a well-behaved function which returns $O(1)$ constant for $x=0$ and decays polynomially as $1/x$ for $x \gg 1$.

The contribution from all larger eigenvalues, $\lambda_k$ with $k \sim O(N)$, to $\chi_{\rm SG}$ can be written in an integral form similar to Eq.~(\ref{another_chi_SG}) by substituting the lower band edge $\lambda_{\rm min}$ with $\lambda_{\rm min}+O(N^{-1/(1+\gamma)})$ 
\begin{eqnarray} 
\nonumber
\frac{1}{N} \sum_{k\sim O(N)} \overline{\lambda_k^{-2}}
& \sim&
 \int_{\lambda_{\rm min}+O(N^{-1/(1+\gamma)})} d\lambda\,\lambda^{-2} \rho(\lambda)
\\
\nonumber
& \propto &
\left(\lambda_{\rm min}+O(N^{-1/(1+\gamma)})\right)^{-(1-\gamma)} 
\\
&\sim&
 t^{-1} h\left(t N^{\frac{1-\gamma}{1+\gamma}}\right)\;. 
\end{eqnarray} 
In the last relation, we have used again the scaling relation for $\lambda_{\rm min}$, and $h(x)$ is another well-behaved function that is proportional to $x$ for $|x| \ll 1$ and converges to a certain constant as $x \to \infty$. Combining the two contributions, we obtain the finite size scaling relation of $\chi_{\rm SG}$ as 
\begin{eqnarray}
\chi_{\rm SG}&\simeq & N^{\frac{1-\gamma}{1+\gamma}}g\left(t N^{\frac{1-\gamma}{1+\gamma}}\right) +t^{-1} h\left(tN^{\frac{1-\gamma}{1+\gamma}}\right) \cr &=&N^{\frac{1-\gamma}{1+\gamma}}F\left(tN^{\frac{1-\gamma}{1+\gamma}}\right)\;, 
\label{finalscaling} 
\end{eqnarray}
where $F(x)\equiv g(x)+h(x)/x$. The properties of $g(x)$ and $h(x)$ guarantee that $F(0)$ is a finite constant and $F(x) \sim 1+O(x^{-1}) $ for $x \gg 1$.

For $H=0$ and $C \ge 3$, $\gamma=1/2$ yields the scaling law $\chi_{\rm SG}=N^{1/3} F(tN^{1/3})$. This relation was often assumed in earlier studies on SG models of the mean field type \cite{Jorg,Billoire2003a,Billoire2003b}. However, as far as the authors know, there has been no numerical validation of this relation, in particular, for the scaling exponent with respect to $t=|T-T_c|/T_c$, even for the case of $H=0$. In addition, there is no theoretical guarantee that $\gamma=1/2$ always holds for $H > 0$ case. As there are only few analytical schemes available for dealing with SG models of finite dimension, we need to build up a solid basis for numerical studies. Circumstantially comparing the results of numerical experiments and theoretical predictions of Eqs.~(\ref{chisg}) and (\ref{finalscaling}) for the current system is a great step towards fulfilling this purpose.

\section{Numerical Experiments}
In order to verify the above-mentioned behavior around the AT instability, we performed large numerical experiments on systems with $C=4$ and sizes $N=2^5,2^6, \ldots ,2^{10}$, 
using the replica exchange (parallel tempering) Markov chain Monte Carlo (MCMC) method \cite{HukushimaNemoto,Hansmann}. Apart from some test runs on small systems with $H=0$, we ran extensive simulations on fields $H=0.1, 0.2, 0.3$ at respectively 33, 34, and 36 different temperatures distributed around $T_c$. For equilibrating the systems, we performed $2^{21}$ MC sweeps (MCSs) and computed thermal averages from $2^{21}$ more MCSs after the equilibration time. Equilibration was tested by comparing the averages obtained by using half and one quarter of the total MCSs. To accelerate equilibration, replicas of adjacent temperatures were exchanged once every 30 MCSs. We simulated 16000 samples for each size.

\begin{figure} 
\begin{center} 
\includegraphics[width=\columnwidth]{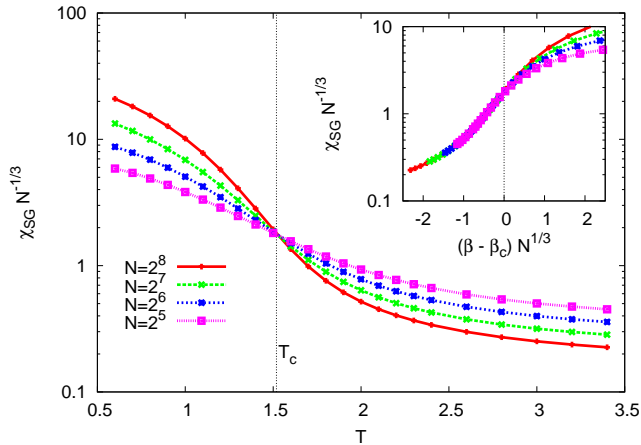} 
\end{center} 
\caption{Rescaled spin glass susceptibility $\chi_{\rm SG}$ without external field.} 
\label{fig0} 
\end{figure}

Figure~\ref{fig0} shows the results of runs with $H=0$. The spin glass susceptibility rescaled by a factor $N^{-1/3}$ (corresponding to $\gamma=1/2$) nicely crosses at the critical temperature, as predicted analytically. The inset should show the scaling function (if finite size effects were absent), but we can clearly see that the data collapse is good only in the high temperature (low $\beta=T^{-1}$) region.

{
Next, let us turn to the case of external fields. Expanding $\left (\left \langle S_i S_j \right \rangle -\left \langle S_i \right \rangle \left \langle S_j \right \rangle \right)^2$ as 
\[ \left \langle S_i S_j \right \rangle \left \langle S_i S_j \right
\rangle -2 \left \langle S_i S_j \right \rangle \left \langle S_i
\right \rangle \left \langle S_j \right \rangle +\left \langle S_i
\right \rangle \left \langle S_j \right \rangle \left \langle S_i
\right \rangle \left \langle S_j \right \rangle
\]
and using different real replicas for computing different thermal averages at the same time, the above equation becomes
\[ \left \langle S_i^1 S_j^1 S_i^2 S_j^2 \right \rangle
-2 \left \langle S_i^1 S_j^1 S_i^2 S_j^3 \right \rangle
+ \left \langle S_i^1 S_j^2 S_i^3 S_j^4 \right \rangle\;,
\]
we can write the spin glass susceptibility
$\chi_{\rm SG}$ as
\begin{eqnarray}
\chi_{\rm SG} &=& N \overline{\left (\left \langle q_{12}^2 \right \rangle
  -2 \left \langle q_{12}q_{13} \right \rangle +\left \langle q_{13} q_{24}
  \right \rangle \right)}\nonumber\\
 &=& N \overline{\left (\left \langle q_{12}^2 \right \rangle
  -2 \left \langle q_{12}q_{13} \right \rangle +\left \langle q_{12} \right \rangle^2 \right)}\;, 
\label{chiSGnumeric} 
\end{eqnarray}
where $q_{ab}=N^{-1} \sum_{i=1}^N S_i^a S_i^b$ is the overlap between
two replicas.  Actually we computed $\chi_{\rm SG}$ in
Eq.(\ref{chiSGnumeric}) by measuring overlaps from four different
replicas $a,b=1,2,3,4$ evolving independently, in order to reduce
correlations effects and the noise-to-signal ratio.
}

\begin{figure} 
\begin{center} 
\includegraphics[width=\columnwidth]{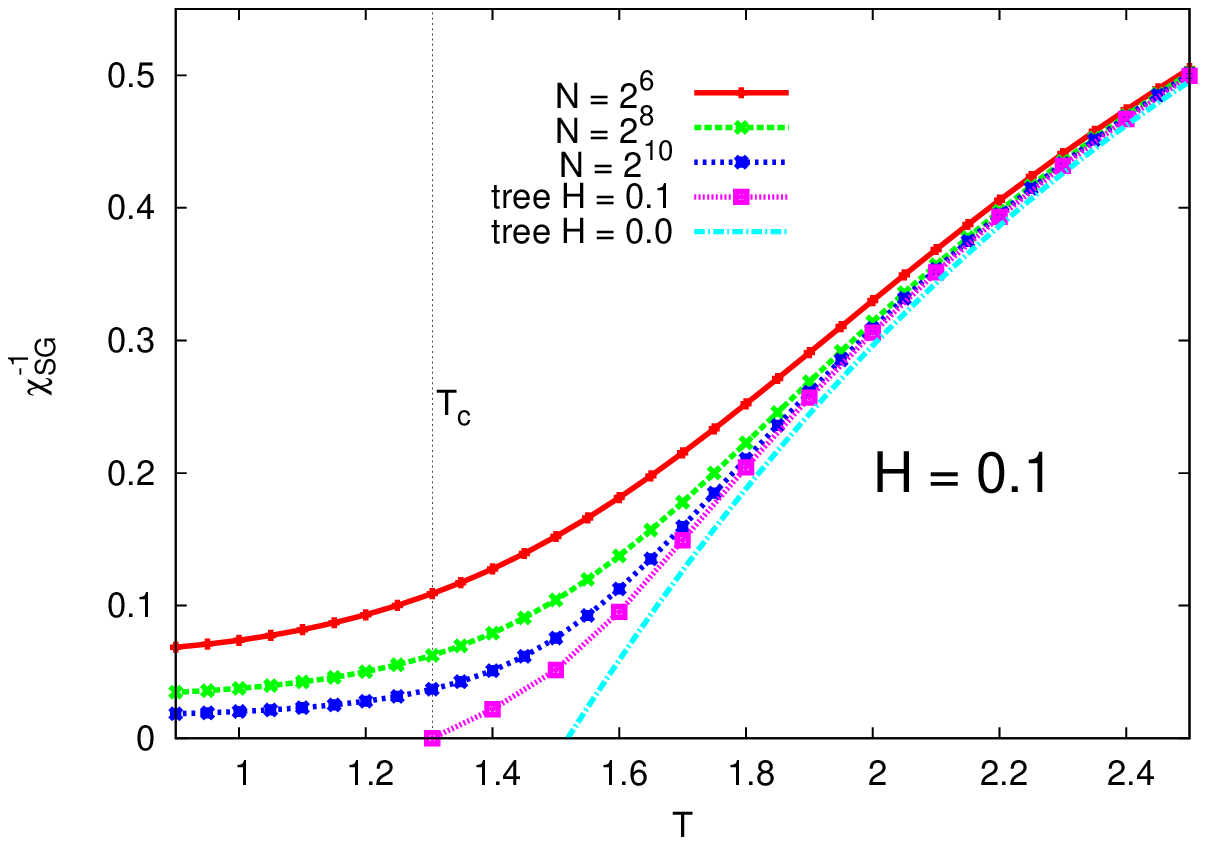} 
\includegraphics[width=\columnwidth]{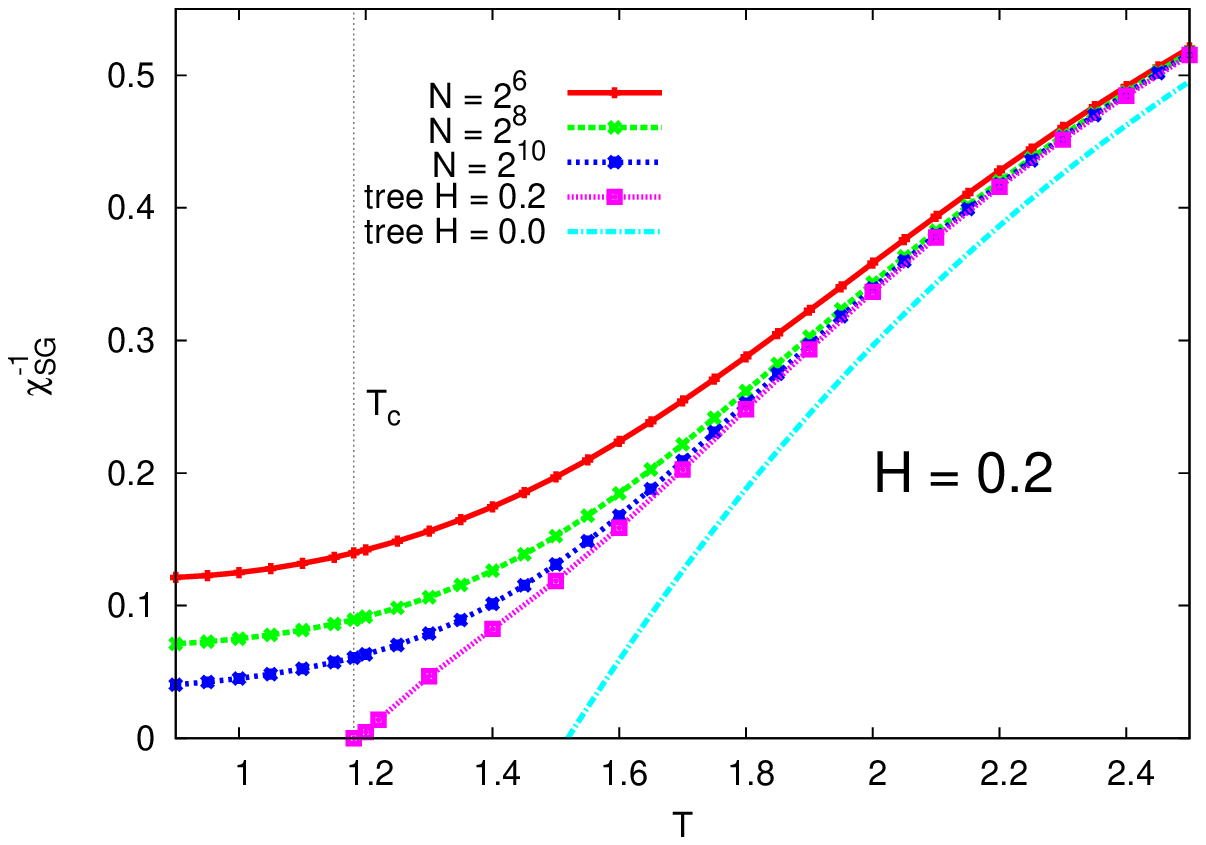} 
\includegraphics[width=\columnwidth]{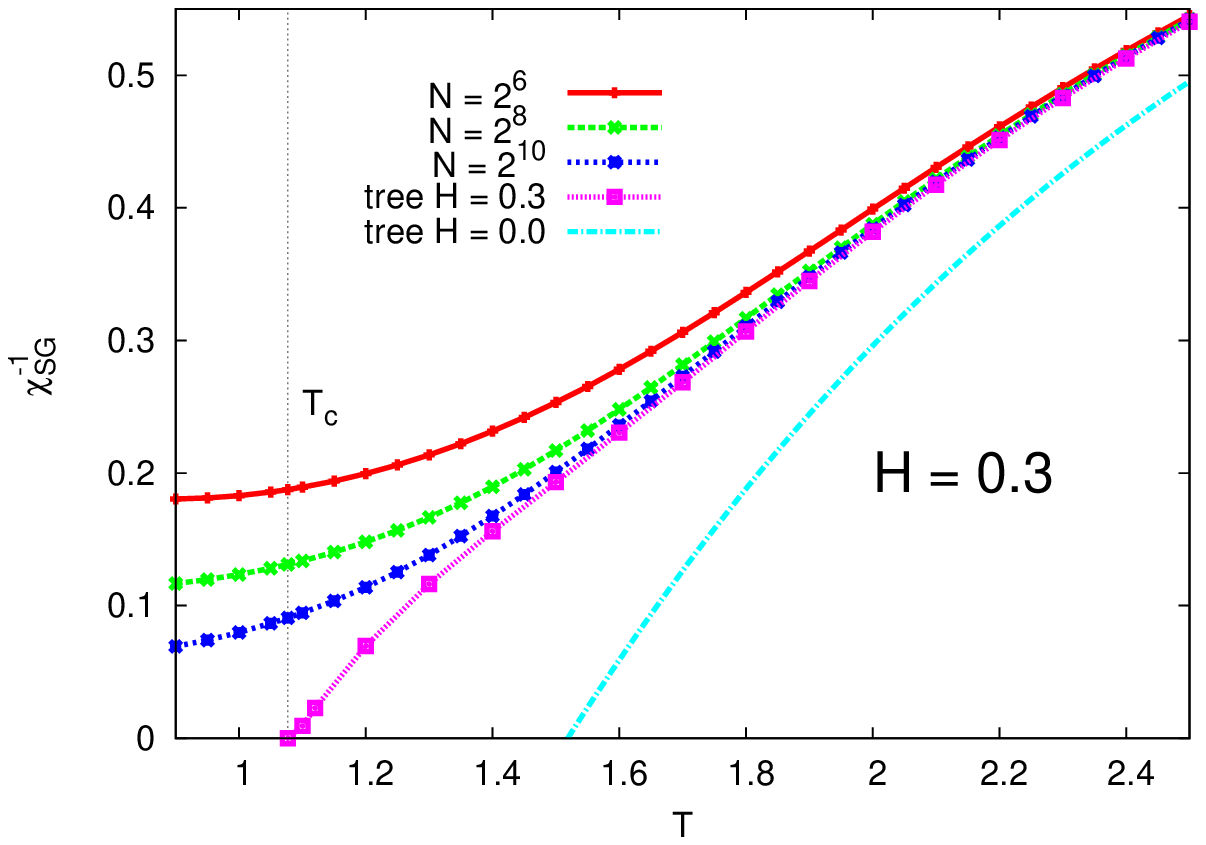} 
\end{center} 
\caption{Inverse of the spin glass susceptibility versus temperature for fields $H=0.1$ (top), $H=0.2$ (middle), and $H=0.3$ (bottom).} 
\label{fig1} 
\end{figure}

Figure~\ref{fig1} compares the susceptibilities $\chi_{\rm SG}$ measured numerically on systems of sizes $N=2^6,2^8,2^{10}$ with the ones computed analytically on the Bethe tree for $H=0.1, 0.2$, and $0.3$. We see that the numerical data at high temperatures converge nicely to the theoretical estimates on the Bethe tree. Note that the diameter of the regular random graph with $C=4$ is only $\ln (N)/\ln(C-1) \simeq 6.3$ even for the case of $N=2^{10}$. 
This indicates that accuracy of the Bethe approximation is not determined only by the size of the graphs or, more precisely, by the length of the shortest loops. The relative strength of the self-interactions compared to the size of the graphs plays a key role in determining the accuracy. In other words, even if the graph contains many loops, which may significantly contribute to self-interaction terms (that are missing on trees), the lack of correlation in the topology makes the net contribution of these loops very small. The final result is that the critical window size for the susceptibilities scales as an inverse power of $N$ rather than $1/\ln N$.

Figure~\ref{fig1} (top) shows the analytical curves corresponding to $H=0$ and $H=0.1$. One can see how the $H=0.1$ data closely follow the $H=0$ curve as long as $T \gtrsim T_c(H=0)$. Only below $T_c(H=0)$ does the data change its curvature and acquire the correct linear behavior in $T-T_c(H=0.1)$. Unfortunately, this change happens at very large values of $\chi_{\rm SG}$, and thus, the asymptotic scaling behavior may be difficult to observe. For larger fields, $H=0.2$ and $H=0.3$, the influence of the $H=0$ fixed point is weaker and the theoretical susceptibility curves are qualitatively similar to the $H=0$ curve, with the linear part in $T-T_c$ extending over a wider range. Nonetheless, for these larger fields, the values of $\chi_{\rm SG}$ are much smaller, and thus, the asymptotic behavior may be difficult to observe in this case as well.

\begin{figure} 
\begin{center} 
\includegraphics[width=\columnwidth]{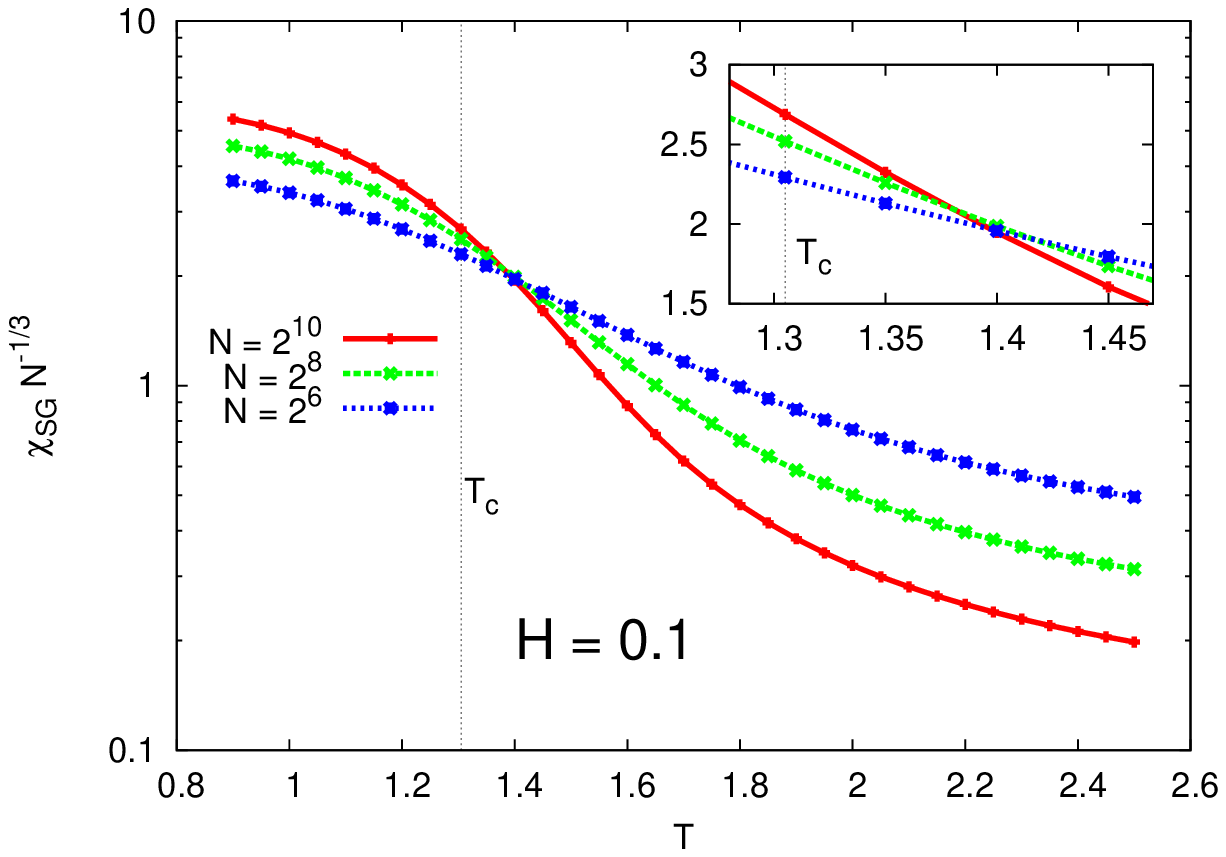} 
\includegraphics[width=\columnwidth]{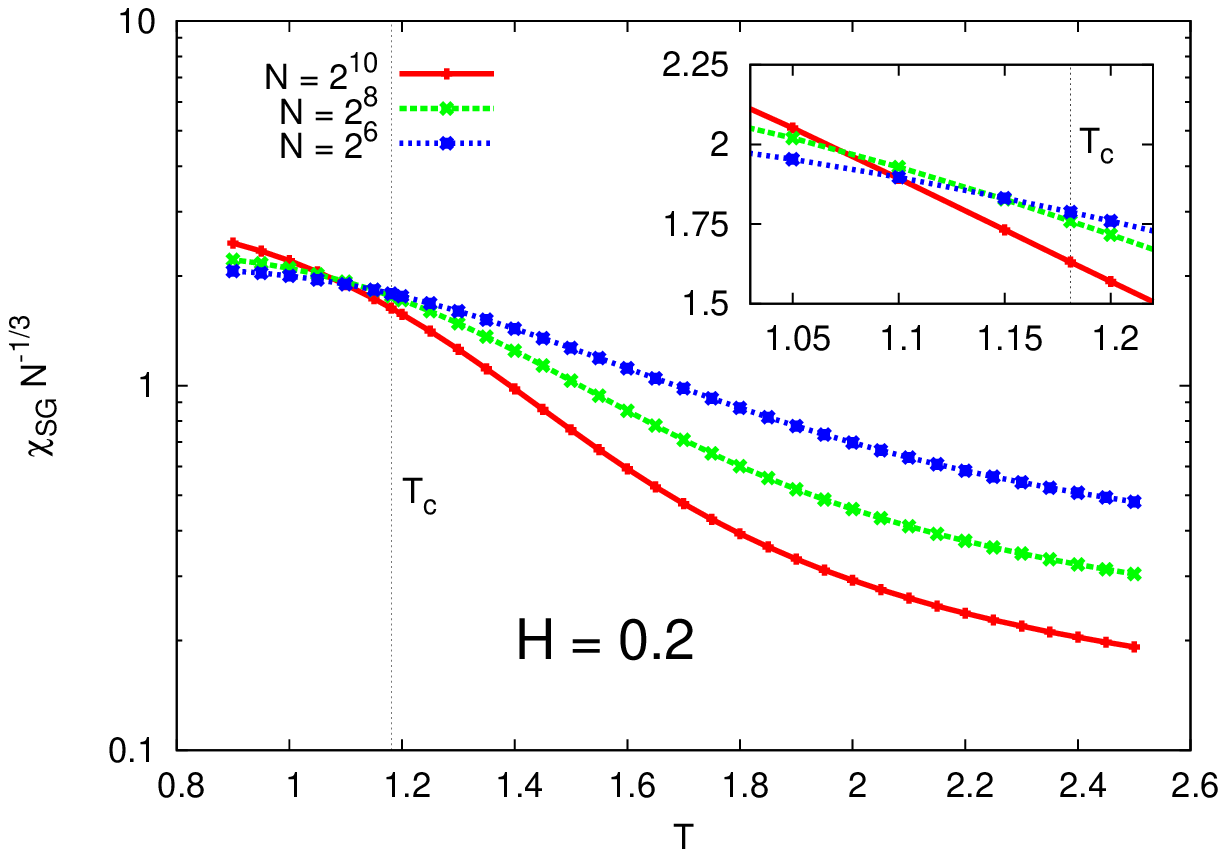} 
\includegraphics[width=\columnwidth]{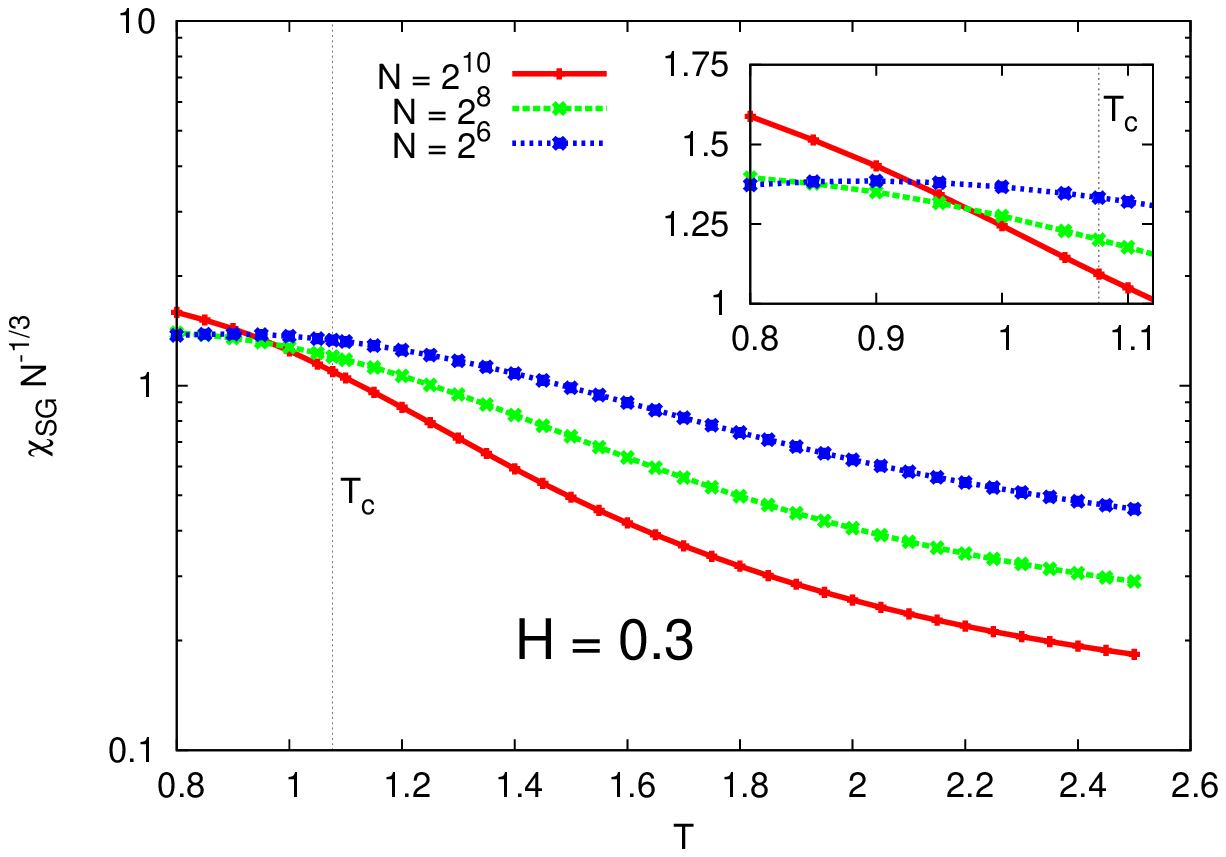} 
\end{center} 
\caption{Rescaled spin glass susceptibility (assuming $\gamma=1/2$) versus temperature for fields $H=0.1$ (top), $H=0.2$ (middle), and $H=0.3$ (bottom). Errors are smaller than the symbol size.} 
\label{fig2} 
\end{figure}

We tried to identify the critical point, $T_c$, from the finite size scaling of the numerical data, by using $\gamma=1/2$ for any field value. For this, we plotted $N^{-1/3} \chi_{\rm SG}$ versus temperature and looked for a crossing point of data sets having different $N$, which should correspond to $T_c$ in the thermodynamic limit. We see from Fig.~\ref{fig2} that finite size corrections are rather large, especially for $H=0.1$ and $H=0.3$, and \emph{change sign depending on the value of the field}. The insets in Fig.~\ref{fig2} zoom in on the region containing the crossings for all data, to reveal whether or not the crossings move towards the analytical $T_c$ value computed under the tree approximation. The $H=0.1$ crossing points move in the right direction, but they do so \emph{very} slowly; most probably due to the $H=0$ fixed point at $T_c(H=0) \simeq 1.52$ in the vicinity. The $H=0.3$ crossing points also move towards $T_c$ and do so faster than those of $H=0.1$, although they come from the low temperature phase. The $H=0.2$ crossing points are more complex, because the crossing point apparently move leftward, away from the critical point. The most natural explanation for this behavior is that for larger system sizes the crossing point moves back again towards $T_c$, as in the case of $H=0.3$ (where the crossing point moves rightward).

Clearly in this model (and probably in any spin glass model in a field), there are finite size effects with opposite signs in competition, whose relative weights depend on the value of the external field. This makes the extrapolation to the thermodynamic limit extremely challenging, especially in cases like the one in Fig. \ref{fig2} (middle) where any reasonable extrapolation would predict a critical temperature (if any) much lower than the true one. This phenomenon must be taken into account when one wants to exclude any phase transition because of lack of a crossing point in the scaled data. See, for example, Refs.~\onlinecite{KLY09,LPRR09} for a recent discussion about this issue for mean-field and non-mean-field spin glasses in an external field.

\begin{figure} 
\begin{center} 
\includegraphics[width=\columnwidth]{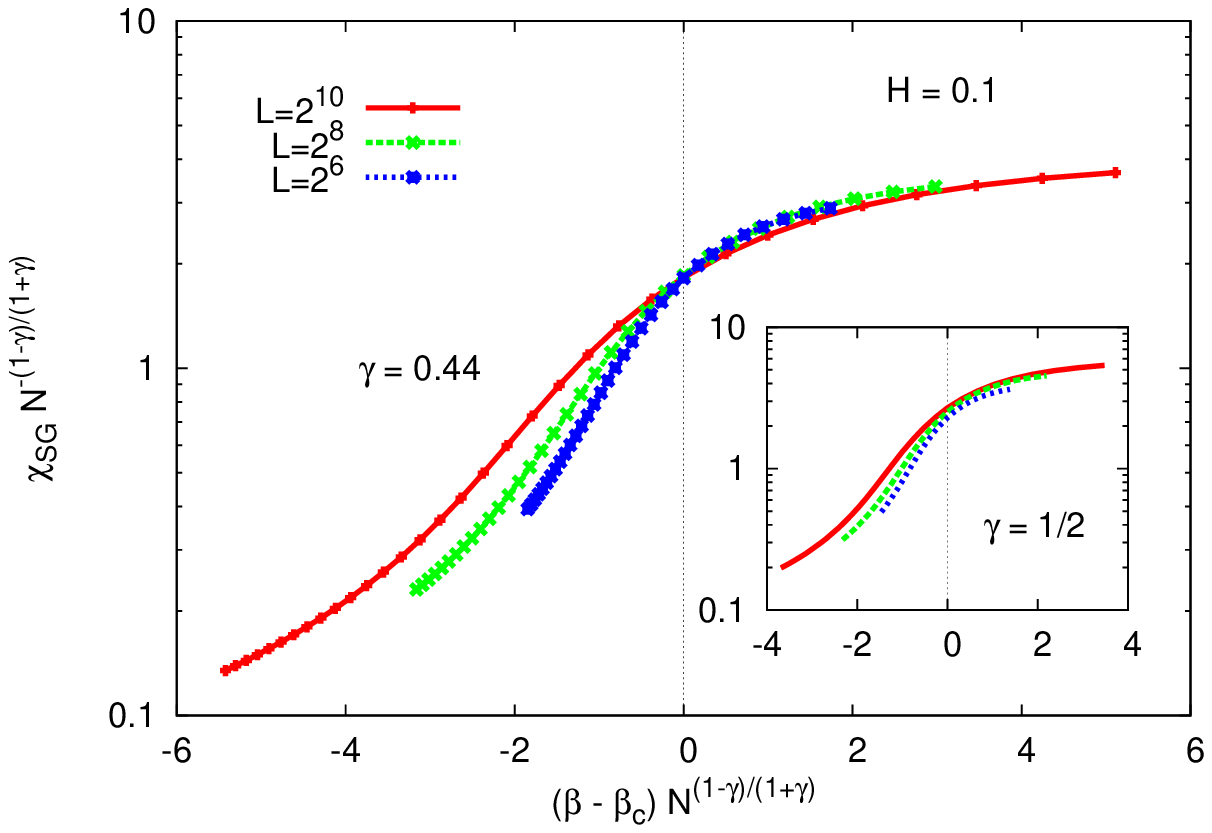} 
\includegraphics[width=\columnwidth]{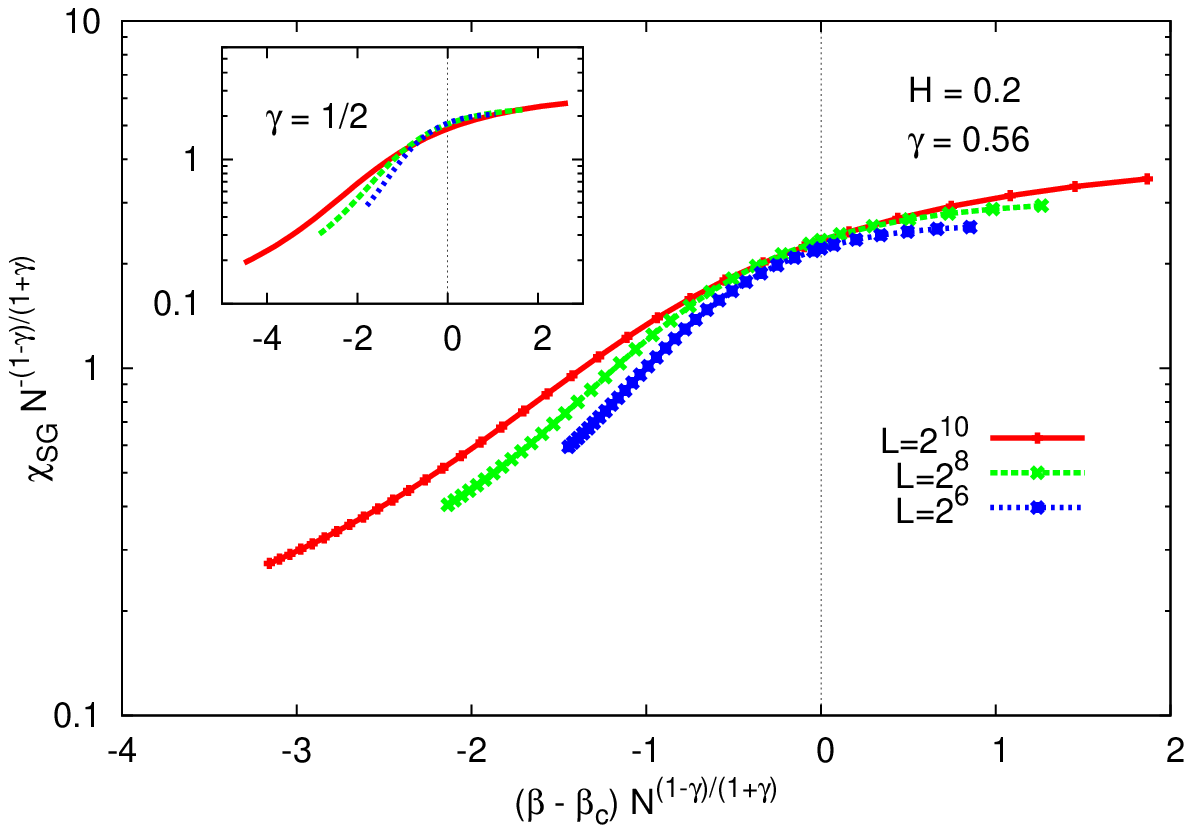} 
\includegraphics[width=\columnwidth]{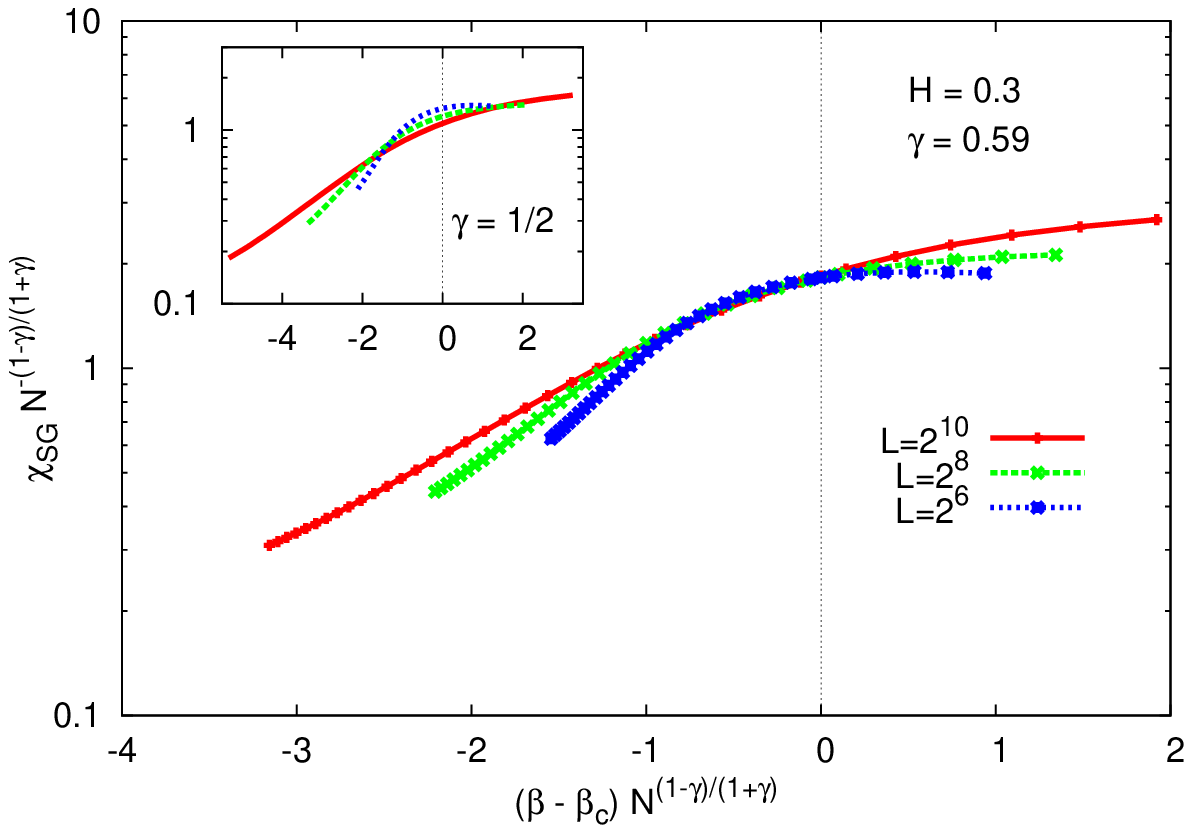} 
\end{center} 
\caption{``Tentative'' scaling function for the spin glass susceptibility for fields $H=0.1$ (top), $H=0.2$ (middle), and $H=0.3$ (bottom). In the main panels, the parameter $\gamma$ has been optimized such as to superimpose the data from the largest sizes at the critical point. In the insets, the value $\gamma=1/2$ is used. Errors are smaller than the symbol size.} 
\label{fig3} 
\end{figure}

{Despite the large finite size effects, we have tried to scale the data
according to Eq.(\ref{finalscaling}).  The results are obviously very
poor and have been reported in Fig.~\ref{fig3} for the sake of
completeness.  Even optimizing over the choice of the $\gamma$
parameter (such as to superimpose the data from largest sizes at the
critical point) we get no data collapse at all (see main panels in
Fig.~\ref{fig3}).  The tentative scaling with $\gamma=1/2$ is shown in
the insets of Fig.~\ref{fig3} and it is even worst.

We should note \textit{en passant} that a much better scaling of the numerical data can be obtained by using two different exponents for the rescalings of the $x$- and $y$-axes. This scaling would imply a divergence of the SG susceptibility in a field as $\chi_{\rm SG} \propto |t|^{-a}$ with $a \simeq 1.6 \div 1.7$. However, given the strong analytical arguments in favor of $a=1$, we have to conclude that such an alternative scaling is only due to finite size effects.}

In order to examine the reason for the poor consistency with the scaling law with $\gamma=1/2$, i.e.\ $\chi_{\rm SG}=N^{1/3} F(t N^{1/3})$, we numerically evaluated the eigenvalues of the Hessian matrix $A=\chi^{-1}$ for systems of $N=2^5, 2^6, 2^7$, and $2^8$. The reduction of the system size is simply due to limited computational resources; the eigenvalue analysis costs much more than the evaluation of $\chi_{\rm SG}$.

\begin{figure} 
\includegraphics[width=78mm]{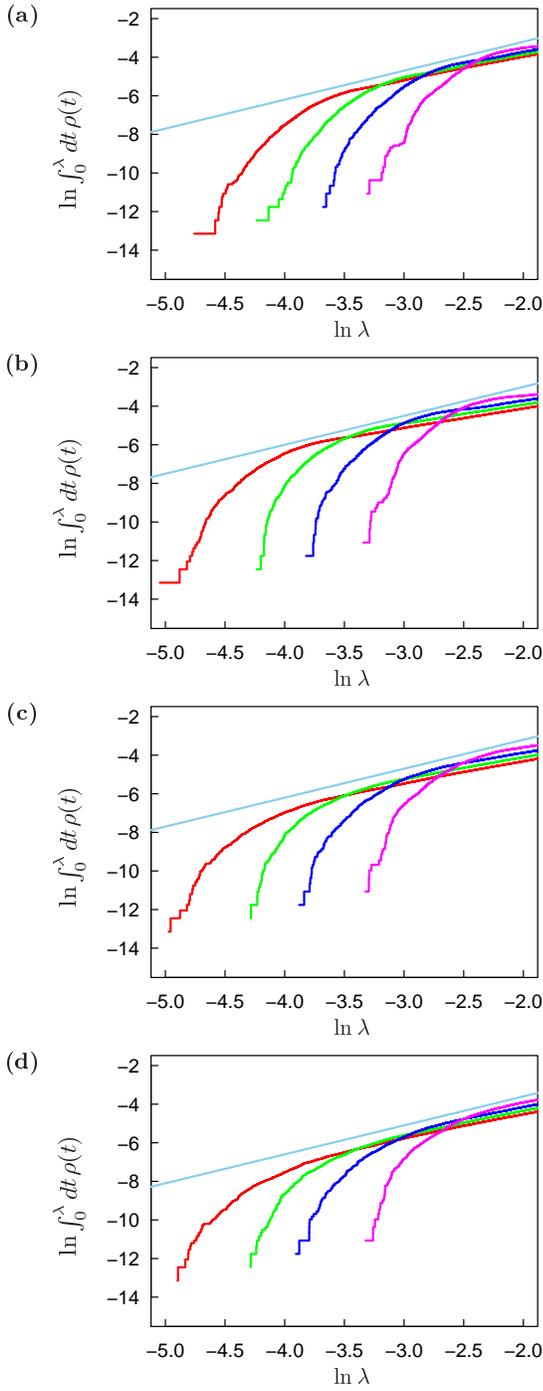}
\caption{Cumulative eigenvalue distribution $\int_{0}^{\lambda} dt \rho(t)$ at the critical temperature $T_c$ for (a): $H=0$, (b): $H=0.1$, (c): $H=0.2$, and (d): $H=0.3$. From right to left, the curves correspond to $N=2^5, 2^6, 2^7$, and $2^8$. The straight lines stand for a scaling relation of $\int_{0}^{\lambda} dt \rho(t) \propto \lambda^{1+\gamma}$ with $\gamma=1/2$.} 
\label{fig6} 
\end{figure}

Figures \ref{fig6} (a)--(d) show the cumulative distributions $\int_0^{\rm \lambda}dt \rho(t)$ at the critical temperature $T_c$, which were numerically computed from data of $2000$ sample systems. These distributions should be proportional to $(\lambda-\lambda_{\rm min})^{1+\gamma} \to \lambda^{1+\gamma}$ in the limit of $N \to \infty$ since $\lambda_{\rm min}$ vanishes at criticality. However, finite size corrections modify the behavior of the numerical data in the vicinity of $\lambda=0$. The envelopes of the cumulative distributions for various $N$ exhibit reasonable consistency with the scaling form of $\gamma=1/2$, which is represented by straight lines in Figs.~\ref{fig6}, for all cases of $H=0, \ 0.1, \ 0.2$, and $0.3$.

\begin{figure}
\includegraphics[width=76mm]{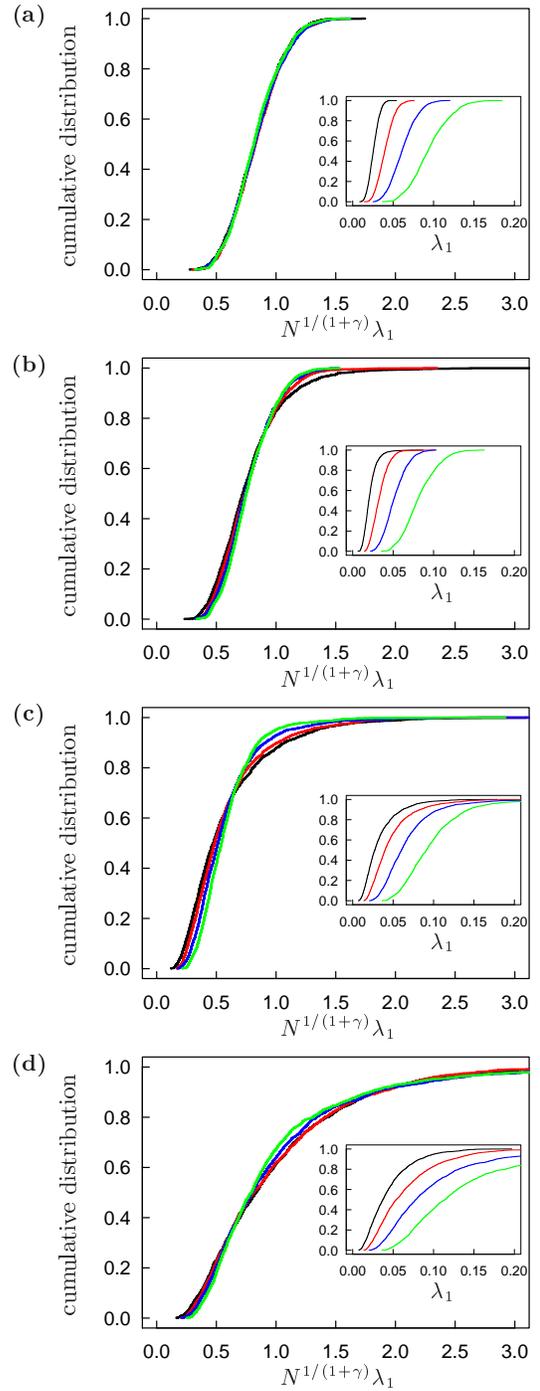} 
\caption{Scaling plots for the cumulative distributions of the smallest eigenvalue $\lambda_1$ at the critical temperatures $T_c$ for (a): $H=0$, (b): $H=0.1$, (c): $H=0.2$, and (d): $H=0.3$. 
{
In each plot, four curves correspond to $ N=2^5, \ 2^6, \ 2^7$, and $2^8 $. 
The plots are obtained by rescaling the horizontal axis of the original ones (insets: $ N=2^5, \ 2^6, \ 2^7$, and $2^8 $
from right to left) 
} as $\lambda_1 \to N^{1/(1+\gamma)} \lambda_1$, where $1/(1+\gamma)$ is determined on the basis of the arithmetic averages of $ \lambda_1 $ over $2000$ samples. The estimates of $\gamma$ are $0.598, \ 0.545, \ 0.943$, and $0.789$ for $H= 0.0, \ 0.1, \ 0.2$, and $0.3$, respectively. Except for (a), the data do not collapse to a single curve with good accuracy.} 
\label{fig7} 
\end{figure}

Another possible source of inconsistency concerning the finite size scaling relation of $\chi_{\rm SG}$ is the finite size correction to the lower band edge $\lambda_{\rm min}$. The argument presented in Sec.~\ref{finiteSize} relies on the assumption that the smallest eigenvalue $\lambda_1$ behaves as $\lambda_1 \simeq \lambda_{\rm min}+N^{-1/(1+\gamma)}\xi_1 \simeq N^{-1/(1+\gamma)}\xi_1$ at $T_c$, where $\xi_1$ is a random variable independent of $N$. This means that the distributions of $\lambda_1$ for different $N$ collapse to a single curve for fixed $H$ after rescaling of $\lambda_1 \to N^{1/(1+\gamma)} \lambda_1$ using an appropriate value of $\gamma$. Figure \ref{fig7} (a) shows that this holds to good accuracy for $H=0$ with $\gamma=0.598$, which is reasonably close to the theoretical prediction $1/2$. However, Figs. \ref{fig7} (b)--(d) and Table \ref{table2} indicate that such a clear scaling relation does not hold for $H > 0$.

\begin{table} 
\begin{tabular}{c|cccc} 
\hline
\hline ~$H$~ &0 & 0.1& 0.2& 0.3 \\
\hline ~$\gamma$ ($\overline{\lambda_1}$)~ & 0.598$\pm$0.012 & 0.545$\pm$0.010 & 0.943$\pm$0.014 & 0.789$\pm$0.015 \\
\hline ~$\gamma $ ($\overline{\lambda_1^2}$)~ &0.599$\pm$0.013 & 0.589$\pm$0.021 & 1.139$\pm$0.001 & 0.730$\pm$0.033 \\
\hline 
\end{tabular} 
\caption[]{Values of $\gamma$ estimated from the arithmetic averages over 2000 samples of $\lambda_1$ ($\overline{\lambda_1}$) and those of $\lambda_1^2$ ($\overline{\lambda_1^2}$). Errors represent the standard errors of the estimates. For $H=0$, the two estimates coincide with each other up to the second digit, justifying a scaling of the form $\lambda_1 =N^{-1/(1+\gamma)} \xi_1$, where $\xi_1$ is a certain random variable that is independent of $N$. On the other hand, they differ significantly for $H =0.1, \ 0.2$, and $0.3$, which implies that a scaling of that form does not hold for $H > 0$.} 
\label{table2} 
\end{table}

{ This is presumably because, in the presence of an external field, the
lower band edge $\lambda_{\rm min}$ has strong 
finite size corrections and a non-negligible dependence on $N$.  
The critical temperature $T_c$ is defined by the condition 
$\lambda_{\rm min} =0$ in the limit of $N \to \infty$, but 
for finite $N$ in the presence of an external field,  
it may not vanish due to a positive bias as $\lambda_{\rm min}=B_1 N^{-\sigma}$
even at $T_c$. 
{
Such a bias could come out if statistical correlations 
among the eigenvalues 
$\lambda_1,\lambda_2, \ldots, \lambda_N$ are 
not negligible. 
}
Consequently, the expression for the smallest eigenvalue
$\lambda_1$ at $T_c$ should be modified
{
at least} as
\[
{\lambda_1} = B_1 N^{-\sigma} + {\xi_1} N^{-1/(1+\gamma)}\;,
\]
where random variable $\xi_1$ generally obeys a certain nontrivial distribution
whose mean is not guaranteed to vanish \cite{TracyWidom}.  Unless
$\sigma$ and $1/(1+\gamma)$ are very different (or very similar), both
scaling terms will be needed for appropriately handling data with $N$
of several hundreds, which are the practical upper limits of the
system size that we can deal with by standard computational resources
to date.  However, estimating the two exponents simultaneously from
data of only few values of $N$, is far from trivial in the
presence of statistical fluctuations.  
}

The presence of strong finite size corrections for $\lambda_{\rm min}$ and the practical difficulty of identifying the scaling relation of the corrections from numerical data mean that accurate numerical evaluation of $T_c$ is very difficult in the presence of external fields. Although we herein examined random sparse networks, a similar issue should also affect other systems. This may be a major reason why the AT line has not been clearly observed in SG models of finite dimensions.

\section{Summary}
In summary, we have explored a finite size scaling relation of the de Almeida-Thouless (AT) instability criticality in random sparse networks analytically and numerically. The spin glass susceptibility $\chi_{\rm SG}$ is a proper measure for signaling the AT criticality. On the basis of the similarity between the random sparse networks and the Bethe trees, we have derived a scheme for evaluating $\chi_{\rm SG}$ of infinitely large systems utilizing the belief propagation algorithm, which makes it possible to evaluate the critical temperature $T_c$ with a numerically feasible procedure. The singularity at $T_c$, however, cannot be directly observed in finite systems due to finite size effects. Therefore, we examined how the finite size scaling relation is determined by the lower band edge behavior of the Hessian matrix.

The validity of the theoretical predictions was examined in extensive numerical experiments. For sufficiently high temperatures, the numerically computed values of $\chi_{\rm SG}$ were reasonably consistent with the theoretical predictions regardless of whether an external field was present. This result supports a widely believed equivalence between random sparse networks and Bethe trees, implying that $T_c$ is the same in both systems. Accordingly, we investigated the consistency between the numerical data and the theoretically obtained finite size scaling relation $\chi_{\rm SG}=N^{\omega}F(N^{\omega}|t|)$. In the absence of an external field, the numerical data are in good agreement with the scaling relation of $\omega=1/3$, as was believed in earlier studies. On the other hand, the consistency of the data with respect to the scaling relation becomes very poor in the presence of a field; the crossing points of $N^{-1/3} \chi_{\rm SG}$ for $N\le 2^{10}$ fluctuate around the theoretical values of $T_c$ non-monotonically with the strength of the field $H$, and the data do not fit the scaling relation even if $\omega$ is tuned. Upon examining the eigenvalues of the Hessian matrix on the basis of numerical simulations for $N \le 2^8$, we found that the lower band edge 
of the 
eigenvalue distribution is very sensitive to $H$ at least for $N$ of several hundreds. This might be a major reason for the inconsistency with the theoretical prediction of finite size scaling.

\begin{figure} 
\begin{center} 
\includegraphics[width=\columnwidth]{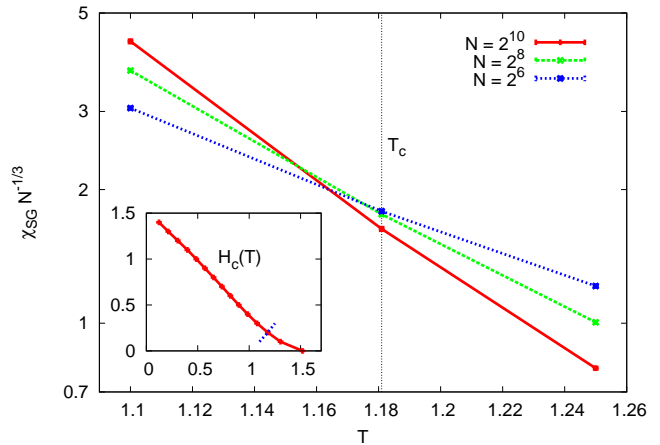}
\end{center} 
\caption{Rescaled spin glass susceptibility (assuming $\gamma=1/2$ as in Fig.~\ref{fig2}) versus temperature. The critical line $T_c(H)$ is crossed perpendicularly. The finite size effects are smaller than in Fig.~\ref{fig2} (middle), but have the same qualitative behavior.} 
\label{fig8} 
\end{figure}

Reference~\onlinecite{Jorg} suggested studying the AT instability along a path in the $(T,H)$ plane that perpendicularly crosses the critical line $T_c(H)$, 
{
which successfully led to accurate estimates of the criticality 
based on data of $N \le 2^9$ in the case of $C=6$}: 
in this way, the finite size effect should be reduced. We followed such a suggestion and analyzed our data along the path represented by the dashed line in the inset of Fig.~\ref{fig8}, which crosses the critical line (full line in the inset of Fig. \ref{fig8}) perpendicularly at $H=0.2$. The resulting susceptibility, scaled by the factor $N^{-1/3}$ as in Fig.~\ref{fig2}, is shown in the main panel of Fig.~\ref{fig8} and should be compared with the data reported in Fig.~\ref{fig2} (middle). It is interesting to note that finite size effects are indeed much reduced (roughly by a factor 4), but the qualitative behavior of the data is exactly the same as in the case with $H$ fixed. In particular, the crossing temperature moves away from the critical temperature, thus giving too small a $T_c$ . Certainly for larger sizes, the crossing temperature will come back to $T_c$ but we have no numerical evidence of that for sizes up to $N=2^{10}$. The same behavior occurs if the data is plotted against the field intensity, as in Fig.~4 of Ref.~\onlinecite{Jorg}.

In light of the present results on the existence of very strong finite
size effects, we believe that the outcome of numerical simulations of
spin glasses in a field should be taken with a lot of care. In
particular, the observation that rescaled susceptibilities do not
cross in a wide temperature range \cite{KLY09} or have a crossing
point moving to low temperatures \cite{Jorg} {
should not
be taken as a \emph{definite} indication for the lack of a spin glass phase.
In the present work we have shown that even in mean field models
 very strong and possibly
non-monotonic corrections to finite-size scaling exist.
In finite dimensional models these corrections are likely to become stronger and
extrapolation from relatively small system sizes is risky.
Maybe the development of new data analysis methods that
may help to reduce finite size effects (like the one in
Ref.~\onlinecite{LPRR09}) would be very welcome.
}

\section*{Acknowledgements} 
The authors would like to thank K. Hukushima, M. Inoue, and E. Marinari for their helpful discussions. 
This study was partially supported by Grant-in-Aids from JSPS/MEXT, Japan, nos. 17340116 and 18079006.

\end{document}